\documentclass{mn2e}
\usepackage{graphicx}
\newcommand{\bq}{\begin{equation}}
\newcommand{\eq}{\end{equation}}
\newcommand{\bqa}{\begin{eqnarray}}
\newcommand{\eqa}{\end{eqnarray}}

\begin{document}

\title[The boundaries of a convective zone \& the extent of overshooting]
{How to define the boundaries of a convective zone and how extended is overshooting?}
\author[L. Deng \& D.R. Xiong]{L. Deng$^1$ \& D.R. Xiong$^2$\\
$^1$National Astronomical Observatories, Chinese Academy of
Sciences, Beijing 100012; licai@bao.ac.cn\\ $^2$Purple Mountain
Observatory, Chinese Academy of Sciences, Nanjing 210008;
Xiongdr@pmo.ac.cn}

\maketitle

\begin{abstract}
Under nonlocal convection theory, convection extends without limit
therefore no apparent boundary can be defined clearly as in the
local theory. From the requirement of a similar structure for both
local and non-local models having the same depth of convection zone,
and taking into account the driving mechanism of turbulent
convection, we argue that a proper definition of the boundary of a
convective zone should be the place where the convective energy flux
(i.e. the correlation of turbulent velocity and temperature) changes
its sign. Therefore, it is convectively unstable region when the
flux is positive, and it is convective overshooting zone when the
flux becomes negative. The physical picture of the overshooting zone
drawn by the usual non-local mixing-length theory is not correct. In
fact, convection is already sub-adiabatic ($\nabla<\nabla_{ad}$) far
before reaching the unstable boundary; while in the overshooting
zone below the convective zone, convection is sub-adiabatic and
super-radiative ($\nabla_{rad}<\nabla<\nabla_{ad}$). The transition
between the adiabatic temperature gradient and the radiative one is
continuous and smooth instead of a sudden switch. In the unstable
zone  the temperature gradient is approaching radiative rather than
going to adiabatic. We would like to claim again that, the
overshooting distance is different for different physical
quantities. In a overshooting zone at deep stellar interiors, the
e-folding lengths of turbulent velocity and temperature are about
0.3H$_P$, whereas that of the velocity-temperature correlation is
much shorter, being about 0.09H$_P$. The overshooting distance in
the context of stellar evolution, measured by the extent of mixing
of stellar matter, should be more extended. It is estimated as large
as 0.25-1.7 H$_p$ depending on the evolutionary timescale. The
larger the overshooting distance, the longer the timescales. This is
due to the participation of extended overshooting tail in the mixing
process.
\end{abstract}


\section{Introduction}

As the classical treatment of convection, the local theory has been
used in modelling stellar structure and evolution. In the
calculation of massive star evolution, Schwarzschild \& H\"arm
(1958) discovered that the hydrogen rich radiative envelope just
outside the helium rich convective core cannot be convectively
stable, and that leaded to the paradox of so called semi-convection.
To solve that problem, the idea of semi-convection was initiated,
i.e. the region outside the convective core is in a state of
semi-convection. Stellar matter in this region is nearly in neutral
stability ($\nabla\leq\nabla_{ad}$), therefore convective energy
transport due to this mild convection can be neglected, while the
mixing of chemical compositions should be important, which makes a
gradient of molecular weight in this region (otherwise called
semi-convection zone). There had been a great debate in the
community for a long period since then on whether the Schwarzschild
or Ledoux criteria should be applied for the neutral stability of
convection, and whether the semi-convective zone should be very wide
or rather narrow. Stothers (1970) commented on various
establishments of semi-convection. Evolutionary scenarios for
massive stars with or without semi-convection were also discussed
(eg. Chiosi \& Summa 1970). It has been realized later that the
problem of semi-convection is in fact due to the non-locality of
stellar convection. Therefore various theories of non-local theory
of stellar convection have been worked out (Spiegel 1963, Ulrich
1970, Xiong 1977, 1981a, 1989, Kufuss 1986, Grossman et al 1993,
Canuto 1993, Canuto \& Dubovikov 1998). Such non-local theories of
stellar convection have then been applied in the studies of
structures of the solar and stellar convective envelopes (Travis \&
Matsushima 1973, Unno et al. 1985, Xiong \& Cheng 1992), stellar
oscillations (Xiong 1981b,  Xiong, Deng \& Cheng 1998, Xiong, Cheng
\& Deng 1998, Xiong \& Deng 2001, 2007) and stellar evolution (Xiong
1986). Generally speaking, non-local theory of convection makes the
results better match observations than the local ones. However, the
non-local theory of convection is rather complicated,
which is much less straightforward to be understood, much more
difficult to be applied and much more computing power demanding than
the phenomenological local (B\"ohm-Vitense 1958) and non-local
mixing length theories (Maeder 1975, Bressan et al. 1981). For these
reasons, it becomes a general practice to use the phenomenological
local or  non-local mixing length treatment for stellar convection
in nowadays stellar evolution models. The non-local mixing of
chemical compositions during the evolution of stars is dealt with by
attaching a parametric overshooting zone outside the convectively
unstable region. The parametric distance of convective overshooting
has a great impact on the properties of stellar evolution. The goal
of present work is to discuss the calibration of the overshooting
distance. The physical definition of the boundary of the convective
zone is discussed in the next section. In section~\ref{sec3},
calibrations of the overshooting distance by numerical simulations
of non-local convection and depletion of solar Lithium
abundance are presented. A summary and discussions are given in the
last section.

\section{How to define the boundaries of a convective
zone?}\label{sec2}

Normally in a local theory of convection, the boundary of convective
zone is given by the so called Schwarzschild's criterion,

\bq \nabla=\nabla_{ad}, \label{eq1}\eq

which is derived by analysis of the local convective stability.
However, if viewed from hydrodynamics strictly, all hydrodynamic
phenomena including convective motions in stars are non-local,
therefore there should be no well defined boundary for convective motion
in an extended medium. In a certain sense, forcing a definition of
boundary for a convective zone is always an artifact. Defining a
boundary for stellar convection is needed in practice of stellar
evolution calculations, but this cannot be done arbitrarily, instead
some objective standards should be respected. These standards should
at least include the following:

\begin{enumerate}

\item As a matter of fact, most of calculations for stellar structure
and evolution still use local theory of convection. Therefore, the
definition of boundary given by a non-local convection theory should
be kept as close as possible to that by a local theory. In other
words, the local and non-local convection models with the same depth
of convective zone should be made to have structures as similar as possbile.

\item the definition of the boundary should be physically pounced
in any case, i.e. the unstable convective zone should be the driving
(excitation) region of convective motion, and the adjacent
overshooting zone should be the dissipation region of convective
motion.

\end{enumerate}

Stellar convection happens due to some internal instability of the
thermal structure in gravitationally stratified fluid, therefore the
study of the resulted convective motions should be based on the
dynamical equations of fluid. A complete dynamic equations of
time-dependent non-local convection theory can be found in our
previous work (Xiong 1981, 1989). For the sake of clarity, and to
make it easier to read this paper, the derivation of the dynamic
equations of turbulent convection in steady fluid is
presented here. The conservations of momentum and energy of fluid
dynamics can be expressed as,

\bqa\lefteqn{\frac{\partial\left(\rho u^i\right)}{\partial t}+
\nabla_k\left(\rho u^iu^k+g^{ik}P\right)}\nonumber\\
 &&\mbox{}+\rho g^{ik}\nabla_k\phi=\nabla_k\sigma^{ik}(u),\label{eq2}
\eqa

\bqa\lefteqn{\frac{\partial\left(\rho H\right)}{\partial t}
+\nabla_k\left(\rho u^kH\right)-\frac{\partial P}{\partial t}}\nonumber\\
 &&\mbox{}-u^k\nabla_kP+\nabla_kF^k_r=\rho\epsilon_N+\sigma^{ik}(u)\nabla_ku_i,
 \label{eq3}
\eqa

where $\rho$, $T$ and $P$ are the regular labels for density,
temperature and pressure (including radiative pressure) of gas; $H$
and $\epsilon_N$ are enthalpy and nuclear energy generation rate per
unit mass; $u^i$ is the $i$th component of fluid motion vector;
$\sigma^{ik}(u)$ the viscous stress tensor; $F^i_r$ is $i$th
component of the radiative flux vector. The implicit summation rule
of tensor calculations is used, i.e. a pair of sub- and super-script
index mains summation from 1 to 3. When convection happens, any
physical quantity can be written as the sum of averaged and
turbulent fluctuated components as,

\bq X=\bar{X}+X'.\label{eq4}\eq

Putting the expressions of all quantities in the form of
eq.~(\ref{eq4}) into eqs.~(\ref{eq2})--(\ref{eq3}), and averaging
the whole equations, the dynamic equations for the mean flow can be
derived as,

\bqa\lefteqn{\frac{\partial\left(\bar{\rho}\bar{u^i}\right)}{\partial
t} +\nabla_k\left(\rho\bar{u^i}\bar{u^k}+\overline{\rho u'^iu'^k}
+g^{ik}\bar{P}\right)}\nonumber\\
 &&\mbox{}=\nabla_k\sigma^{ik}(\bar{u}),\label{eq5}\eqa

\bqa\lefteqn{\frac{\partial\left(\bar{\rho}\bar{H}\right)}{\partial
t}+\nabla_k\left(\bar{\rho}\bar{u^k}\bar{H}+\overline{\rho u'^kH'}\right)}\nonumber\\
 &&\mbox{} -\frac{\partial\bar{P}}{\partial t}
 -\bar{u^k}\nabla_k\bar{P}-\overline{u'^k\nabla_kP}+\nabla_k\bar{F^k_r}\nonumber\\
 &&\mbox{}=\bar{\rho}\bar{\epsilon_N}+\overline{\sigma^{ik}(u)\nabla_ku_i}.
\label{eq6}\eqa

Subtracting the corresponding equations of the mean motion
eqs.~(\ref{eq5})--(\ref{eq6}) from eqs.~(\ref{eq2})--(\ref{eq3}),
and considering the static state of the flow, i.e. \bq
\bar{u}=\frac{\partial\bar{P}}{\partial
t}=\frac{\partial\bar{\rho}}{\partial
t}=\frac{\partial\bar{H}}{\partial t}=0,\label{eq7}\eq

the dynamic equations for the fluctuation quantities can be derived
as,

\bqa\lefteqn{\frac{\partial w'^k}{\partial t}
+{1\over\bar{\rho}}\left(g^{ik}P'+\rho u'^iu'^k-\overline{\rho u'^iu'^k}\right)}\nonumber\\
 &&\mbox{} +g^{ik}\left[\frac{\rho'}{\bar{\rho}}\nabla_k\bar{\Phi}+\nabla_k\Phi'\right]
 ={1\over\bar{\rho}}\nabla_k\sigma^{ik}(u'),\label{eq8}\eqa

\bqa\lefteqn{{\partial\over{\partial t}}\left(\rho H'+\rho'\bar{H}\right)
 +\nabla_k\left(\rho u'^k\bar{H'}-\overline{\rho u'^kH}\right)}\nonumber\\
 &&\mbox{} -\frac{\partial P'}{\partial t}-u'^k\nabla_kP+\overline{u'^k\nabla
 P}\nonumber\\
 &&\mbox{}
 =\rho\epsilon'_N+\rho'\bar{\epsilon_N}-\nabla_kF'^k_r\nonumber\\
 &&\mbox{}
 +\sigma^{ik}(u')\nabla_ku'_i-\overline{\sigma^{ik}\nabla_ku_i}.\label{eq9}
\eqa

By using a certain thermodynamic relations, and following some
deductions and simplifications, eq.~(\ref{eq9}) can be written
as,

\bqa\lefteqn{\frac{\partial}{\partial t}\left(\frac{T'}{\bar{T}}\right)
 +w'^k\left(\nabla_k\ln\bar{T}-\nabla_{ad}\nabla_k\ln\bar{P}\right)}\nonumber\\
 &&\mbox{} +\frac{1}{\bar{\rho}\bar{c_P}\bar{T}}\left\{u'^k\nabla_kP
 +\nabla_k\left[\bar{\rho}\bar{c_P}\bar{T}\left(w'^k\frac{T'}{\bar{T}}
 -\overline{w'^k\frac{T'}{\bar{T}}}\right)\right]\right\}\nonumber\\
 &&\mbox{} =\frac{1}{\bar{\rho}\bar{c_P}\bar{T}}\left[\nabla_kF'^k_r
 +\sigma^{ik}(u)\nabla_ku_i\right],\label{eq10}
\eqa

where $w'$ is the density weighted fluctuation of turbulent
velocity,

\bq w'^k=\frac{\rho u'^k}{\bar{\rho}},\label{eq11}\eq

Starting from eq.~(\ref{eq8})
and eq.~(\ref{eq10}), we can have the following dynamic equations for
the auto- and cross-correlations of turbulent velocity and
temperature fluctuations:

\bqa\lefteqn{{3\over 2}\bar{\rho}\frac{\partial x^2}{\partial t}
 =B\frac{GM_r\bar{\rho}}{r^2}V}\nonumber\\
 &&\mbox{} +\bar{\rho}\frac{\partial}{\partial M_r}
 \left(4\pi r^2\bar{\rho}\overline{u'_rw'_iw'^i/2}\right)
 -1.56\frac{GM_r\bar{\rho}^2 x^3}{c_1r^2\bar{P} },\label{eq12}
\eqa

\bqa\lefteqn{\frac{\partial Z}{\partial t}=2\frac{GM_r\bar{\rho}}{r^2\bar{P}}
 \left(\nabla-\nabla_{ad}\right)V}\nonumber\\
 &&\mbox{} +\frac{1}{\bar{\rho}\bar{C_P}^2}\frac{\partial}{\partial M_r}
 \left[4\pi r^2\bar{\rho}^2\bar{C_P}^2
 \overline{u'_r\left(\frac{T'}{\bar{T}}\right)^2}\right]\nonumber\\
 &&\mbox{} -1.56\frac{GM_r\bar{\rho}}{c_1r^2\bar{P}}\left(x+x_c\right)Z,\label{eq13}
\eqa

\bqa\lefteqn{\frac{\partial V}{\partial t}=\frac{GM_r\bar{\rho}}{r^2\bar{P}}
 \left(\nabla-\nabla_{ad}\right)x^2+B\frac{GM_r}{r^2}Z}\nonumber\\
 &&\mbox{} +{1\over{\bar{C_P}}}\frac{\partial}{\partial M_r}
 \left(4\pi
 r^2\bar{\rho}\bar{C_P}\overline{u'_rw'_r\frac{T'}{\bar{T}}}\right)\nonumber\\
 &&\mbox{}
 -0.78\frac{GM_r\bar{\rho}}{c_1r^2\bar{P}}\left(3x+x_c\right)V,\label{eq14}
\eqa

where $x^2$, $Z$ and $V$ are respectively the auto and cross
correlations of turbulent velocity $w'$ and the relative temperature
fluctuation $T'/\bar{T}$, defined as the following,

\bq x^2=\overline{w'^iw'_i}/3,\label{eq15}\eq

\bq Z=\overline{\left(T'/\bar{T}\right)^2},\label{eq16}\eq

\bq V=\overline{w'_rT'/\bar{T}},\label{eq17}\eq

while $\nabla_{ad}$ is the adiabatic temperature gradient,
$\nabla=\partial\ln\bar{T}/\partial\ln\bar{P}$ is the temperature
gradient, $x_c$ is a variable related with the effect of thermal
conductivity:

\bq
x_c=\frac{3acGM_r\bar{T}^3}{c_1\bar{\rho}\bar{C_P}\bar{P}r^3}.\label{eq18}
\eq

$P_e=x/x_c$ is the effective Peclet number of turbulent convection.
$\bar{B}=-\left(\partial\ln\rho/\partial\ln T\right)_P$ the expansion
coefficient of gas. The detailed derivation of the dynamic equations
of correlations can be found in our previous work (Xiong 1978, 1981,
1989). Eqs.~(\ref{eq12}),(\ref{eq13}) and (\ref{eq14}) are the
dynamic equations of turbulent convection in steady fluid, which have very clear physical
meanings. Eq.~(\ref{eq12}), for instance, is for the conservation of
turbulent kinetic energy. The left hand side is the rate of
variations of turbulent energy per unit volume, which is equal to
the sum of the 3 terms on the right hand side: the first term is the
work done by buoyant force:

\bq W_{buo}=\frac{GM_r\bar{\rho}}{r^2}BV.\label{eq19}\eq

while the expression in the bracket of the second term represents
the effect of non-locality, which is the flux of turbulent kinetic energy $L_t$,

\bq L_t=4\pi
r^2\bar{\rho}\overline{u'_rw'_rw'^i/2}.\label{eq20}\eq

Therefore the second term on the right hand side of eq.~(\ref{eq12})
is the net gaining rate of turbulent kinetic energy per unit volume
$-\bar{\rho}\partial L_r/\partial M_r$. The third term on the right
hand side of eq.~(\ref{eq12}) is for viscous dissipations, i.e. the
dissipation rate $\bar{\rho}\bar{\epsilon_1}$ by converting
turbulent kinetic energy into thermal energy due to viscosity. $c_1$
is a convection parameter related to the viscous dissipation of
turbulent convection. $l_e=c_1H_Pr/R_0$ is the linear size of the
energy-containing eddies (Xiong 1978, 1981a, 1989). Both
Eqs.~(\ref{eq13}) and (\ref{eq14}) have similar physical means: the
left hand sides are the variation rates of correlations $Z$ (or $V$)
respectively. The first (and the second for eq.~[\ref{eq14}]) term
on the right hand side of eq.~(\ref{eq13}) [ or eq.~(\ref{eq14})] is
the rate of increase of $Z$ (or $V$) due to super-adiabatic
temperature gradient (and buoyant force for eq.~[\ref{eq14}]); the
second (the third for eq.~[\ref{eq14}]) term represents the net
increase rate of $Z$ (or $V$) due to non-local convective energy
flux; while the last term describes the turbulent dissipation due to
viscosity and thermal conductivity. For static convection, the time
derivatives terms on the left hand side of
eqs.~(\ref{eq12})--(\ref{eq14}) all vanish:

\bq \frac{\partial x^2}{\partial t}=\frac{\partial Z}{\partial
t}=\frac{\partial V}{\partial t}=0.\label{eq21}\eq

Hence, eq.~(\ref{eq12}) can be rewritten as,

\bq W_{buo}-\bar{\rho}\frac{\partial L_r}{\partial
M_r}-\bar{\rho}\bar{\epsilon_1}=0.\label{eq22}\eq

It is clear from eq.~(\ref{eq22}) that, for static convection, the
net gaining rate of turbulent kinetic energy (-$\bar{\rho}\partial
L_r/\partial M_r$) and buoyant force work ($W_{buo}$) will be
balanced by turbulent dissipation $\bar{\rho}\bar{\epsilon_1}$ (the
sum of all the three terms vanishes). There is no need to go through
the similar physical meanings of eqs.~(\ref{eq13}) and (\ref{eq14}).
When neglecting all the third order correlations representing the
transportation effect of non-local turbulent convection,
eqs.~(\ref{eq12})--(\ref{eq14}) can be converted to,

\bq BV-1.56\frac{\bar{\rho}}{c_1\bar{P}}x^3=0,\label{eq23}\eq

\bq
\left(\nabla-\nabla_{ad}\right)V
-\frac{0.78}{c_1}\left(x+x_c\right)Z=0,\label{eq24}\eq

\bq x^2\left(\nabla-\nabla_{ad}\right)+B\frac{\bar{P}}{\bar{\rho}}Z-
\frac{0.78}{c_1}\left(3x+x_c\right)V=0,\label{eq25}\eq

It is rather easy to solve eqs.~(\ref{eq23})--(\ref{eq25}) for
$x^2$, $Z$ and $V$,

\bq x^2={1\over
2}\left(\frac{c_1}{0.78}\right)^2\frac{B\bar{P}}{\bar{\rho}}
\left(1+{x\over x_c}\right)^{-1}\left(\nabla-\nabla_{ad}\right),
\label{eq26}\eq

\bq Z=\left(\frac{c_1}{0.78}\right)^2\left(1+{x\over
x_c}\right)^{-2}\left(\nabla-\nabla_{ad}\right)^2,\label{eq27}\eq

\bq
V=\left(\frac{c_1}{0.78}\right)^2\left(\frac{B\bar{P}}{2\bar{\rho}}\right)^{1/2}\left(1+{x\over
x_c}\right)^{-3/2}\left(\nabla-\nabla_{ad}\right)^{3/2}.\label{eq28}\eq

Comparing eqs. (\ref{eq26})-(\ref{eq28}) and the equations of the
local mixing length theory (B\"hom-Vitense 1958), and substituting
$c_1$ and the effective Peclet number $x/x_c$ in our equations by
the mixing length parameter $\alpha$ and $\gamma$ respectively, it
is clear that the two expressions are the same. The stability
condition for convection is the Schwarzschild criterion (eq.~[1]) in
a chemically homogeneous medium. When there is a molecular weight
gradient, the neutral stability condition should be Ledoux criterion
(Xiong 1981a):

\bq\nabla=\nabla_{ad}+\nabla_\mu,\label{eq29}\eq

Hence, viewed from hydrodynamics, the local mixing length theory is
only a special simplified case of our statistical theory of
correlations for turbulent convection. The third order correlation
terms in eqs.~(\ref{eq12})--(\ref{eq14}) represent non-local effect
of turbulent convection, neglecting which makes the equations
becoming the local expressions eqs.~(\ref{eq23})--(\ref{eq25}) or
their explicit form eqs.~(\ref{eq26})--(\ref{eq28}). In this case, a
convective zone will have a clearly defined boundary given by
Schwarzschild (or Ledoux) criterion: $\nabla >\nabla_{ad}$ is
convectively unstable, while $\nabla <\nabla_{ad}$ stable
(radiative). As we shall show later, within the convectively
unstable ($V>0$) zone far away from the boundary, the third order
correlation terms in eqs.~(\ref{eq12})--(\ref{eq14}) can be safely
neglected compared with other terms. This means that the local
expression of convection is a fairly good first approximation at
the deep interior of an unstable zone. This is exactly the reason
why the mixing length theory is still widely applied in the
calculations of stellar structures. However, when studying the
entirety of a convective zone, especially near the boundary of
convective zone and in the overshooting region, the third
correlation terms cannot be neglected. Instead, they are the true
reasons for the existence of convective overshooting. In the
non-local convection theory, it is clear from
eqs.~(\ref{eq12})--(\ref{eq14}) that the turbulent velocity and
temperature fluctuations are different from zero everywhere,

\bq x>0;\,\,\,Z>0.\label{eq30}\eq

These conditions make it difficult and uncertain to define a
boundary of a convective zone. Schwarzschild criterion is
overwhelmingly used in the community to fix the boundary. They think
that the temperature gradient is very near and slightly higher than
the adiabatic temperature gradient in the unstable region; and is
also near but slightly lower than the adiabatic one within the
overshooting zone. We are going to show that such a picture for
convective overshooting is not correct following the dynamical
theory of turbulent convection. The cause of such a mistake is that
there is a implicit hypothesis that has been applied in the
phenomenological mixing length theory: turbulent velocity is fully
correlated (either positively or inversely) with temperature (see
eg. Xiong \& Cheng 1992, Petrovay \& Marik 1995). In fact, when
convection is very effective (the effective Peclet number $x/x_c\gg
1$), the correlation between turbulent velocity and temperature
fluctuations near the boundary and in the overshooting zone
decreases very quickly and vanishes eventually (Xiong \& Cheng
1992).

Eqs.~(\ref{eq12})--(\ref{eq14}) are derived under rather general
conditions among which are two important assumptions as the following:

\begin{enumerate}

\item convection is subsonic, the relative fluctuations of
temperature and density are both far less than unity:

\bq \left|\rho'/\bar{\rho}\right|\ll
1;\,\,\,\left|T'/\bar{T}\right|\ll 1, \label{eq31}\eq

\item inelastic approximation which actually filters out all
acoustic waves not important for energy transfer in subsonic
convection.

\end{enumerate}

It is well known that the dynamic equations of turbulent
correlations have no closure due to the nonlinearity of
hydrodynamics. That means: the third order correlations must be
present in the dynamic equations of the second order correlations;
while the fourth order ones bound to turn up in the equations of the
third order ones, and so forth. Some hypothesis must be used in
order to make a closure for the dynamic equations of the
correlations. Obviously, the closure cannot be unique. Quite a few
methods have been adopted so far (Xiong 1981a, 1989a, Canuto 1993,
Grossman et al. 1993, Canuto \& Dubvikov 1998). In our opinion, a
good closure should meet the following conditions:

\begin{enumerate}

\item the solutions of the resulted equations must be physically
sensible. For instance, the standard quasi-normal approximation
seems to be better in terms of mathematics, however it gives
solutions like $x^2<0$ or $Z<0$ which are physically non-sense
(Grossman 1996). Such a seemingly reasonable assumption, if not
modified somehow, cannot be used for the closure of the dynamical
equations of the third order correlations;

\item the solutions presented should not be in contradiction with
observations. For instance, it should reproduce the main
observational properties of solar granular velocity field, it should
be able to explain the pulsation instabilities of low temperature
stars having extended convective envelope, and it should be able to
model the observed Lithium abundance patterns in the atmospheres of
the Sun and solar type stars, and so on;

\item the solutions provided should be comparable to that of direct
hydrodynamical simulations.

\end{enumerate}

Our non-local theory of convection (Xiong 1981a, 1989a) have been
tested against the above standards, quite satisfactory results have
been reached, therefore we have good reason to believe that it has
nicely expressed the dynamic behaviors of stellar turbulent
convection.

\begin{figure}
\includegraphics[width=84mm]{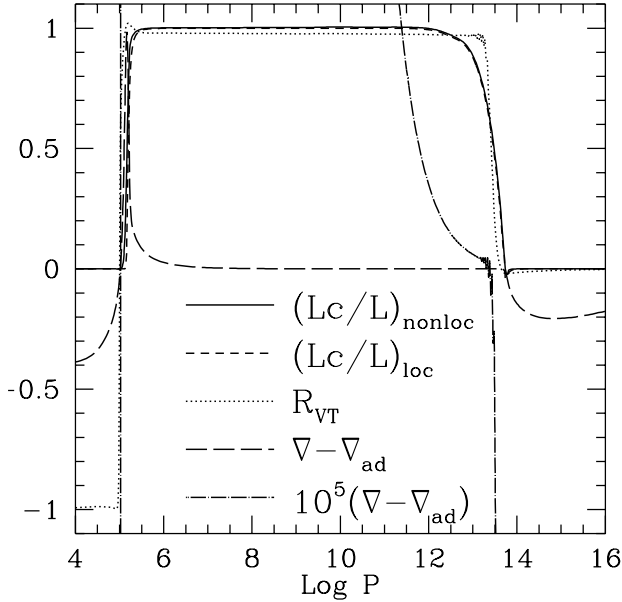}
\caption{The super-adiabatic temperature gradient
$\nabla-\nabla_{ad}$, turbulent velocity-temperature correlation
$R_{VT}$, and the fractional convective flux $L_c/L$ versus the
depth ($\log P$) for a non-local convection model of the Sun. The
dashed line is the fractional convective flux $L_c/L$ for a local
model with the same depth of convective zone.}\label{fig1}
\end{figure}

\begin{figure}
\includegraphics[width=84mm]{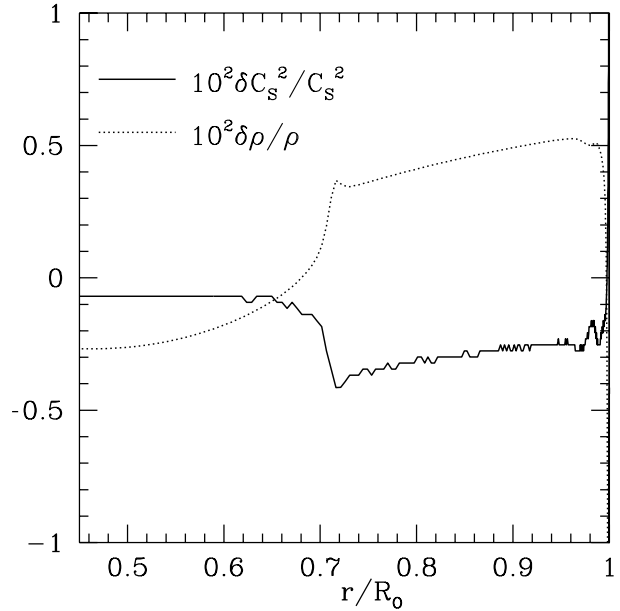}
\caption{The relative differences in the squared sound speed and
density between the non-local and the local convection models with
the same depth of the convective zone versus the fractional
radius.}\label{fig2}
\end{figure}

The solid line in fig.~\ref{fig1} shows the fractional convective
flux $L_c/L$ versus depth $\log P$ for a model of the solar convective
zone calculated with our non-local convection theory, and the dashed
line is that of a local convection model having the same depth of
convective zone. It follows from the figure that there is almost no
difference between the local and non-local model, except some
sizable deviations near the boundary of the convective zone;
Fig.~\ref{fig2} depicts the relative squared sound speed and density
difference between the local and non-local solar convection zone
models with the same depth of convective zone versus depth. The
relative difference between the two models is mostly below 1\%
excluding the solar surface region. In this case, the boundary of
the (non-local) convective zone is set at where the turbulent
velocity-temperature correlation vanishes,

\bq V=0;\label{eq32}\eq

Passing through the boundary, $V$ changes its sign: within the
convective zone:

\bq V>0,\label{eq33}\eq

and in the overshooting zone:

\bq V<0.\label{eq34}\eq

\begin{figure*}
\includegraphics[width=168mm]{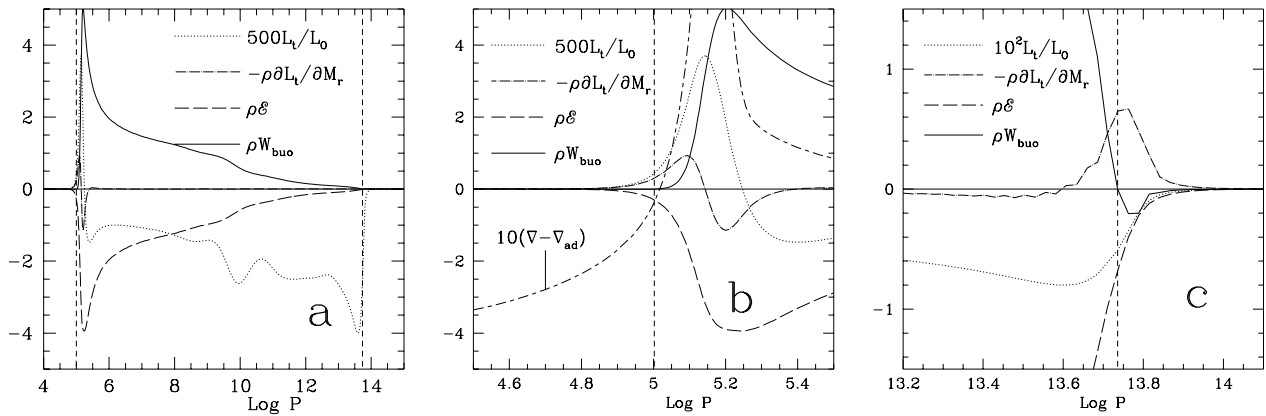}
\caption{a). the work done by buoyant force, viscous dissipation of
turbulence $\rho\epsilon$, fractional turbulent kinetic energy flux
$L_t/L$ and the net gaining rate of kinetic energy due to turbulent
diffusion $-\rho\partial L_t/\partial M_r$ versus depth ($\log P$)
for a non-local solar model. b). and c). are the expanded plots near
the upper and lower boundaries of convective zone indicated by the
vertical dashed lines.}\label{fig3}
\end{figure*}

It is clear from Figs.~\ref{fig1} and \ref{fig2} that if the
boundary is defined as such, the structures of the local and
non-local convection models with the same depth of convective zone
should be similar. It can be understood by the fact that, within the
stellar interior, the turbulent kinetic energy flux ($L_t$) is
generally much less than that of thermal convection ($L_c$), and the
turbulent pressure ($P_t=\rho x^2$) is much less than that of gas
($P_g$). It can be easily shown that $P_t/P_g\sim x^2/C_s^2=Ma^2$,
where $C_s$ is the local sound speed, $Ma$ is the Mach number of
turbulence; $L_t/L_c$ is of the same order of magnitude as that of
$P_t/P_g$. Except at the top of convective zone, we have $Ma\ll 1$.
It follows from fig.~\ref{fig3} that, for the Sun, $L_t/L< 1\%$,
therefore the thermal convection $L_c$ dominates the
pressure-temperature (P-T) structure. It is then clear that, when
defining the boundary of convective zone by $V$ changing its sign,
the structures of the local and non-local models having the same depth of convective
zone will be similar.

Fig.~\ref{fig1} clearly demonstrates that convective motions near
both upper and lower boundaries of the convective zone are very
different. This is due to the fact that, in the atmosphere, the
density is very low and $P_e=x/x_c<1$, therefore convective energy
transfer is inefficient. As a result, there exists a thin
super-adiabatic layer atop of the convective zone. Passing through
the upper boundary, the turbulent velocity-temperature correlation
$R_{VT}=V/xZ^{1/2}$ drops quickly from $\sim 1$ to $\sim -1$. This
theoretical prediction agrees the observations of solar granular
velocity field (Leighton et al.1962; Salucci et al. 1994) and the
results of hydrodynamic simulations (Kupka 2003). Contrary to the
situations in the solar atmosphere, convection is highly efficient
in terms of energy transfer ($P_e\gg 1$) in the deep interiors of
the Sun. Towards the lower boundary of the convective zone, the
turbulent velocity-temperature correlation $R_{VT}$ decreases
abruptly and approaches zero ($\left|R_{VT}\right|\ll 1$). What
makes it so different at the two boundaries is the distinct the
effective Peclet number.

Fig.~\ref{fig3}a shows the work done by buoyant force $\rho
W_{buo}$, the net gain ($>0$) or loss ($<0$) of turbulent kinetic
energy due to non-localism of turbulence $-\rho\partial L_r/\partial
M_r$ and turbulent viscous dissipation rate $\rho\epsilon$ as
functions of depth in the solar convective zone. As shown in
fig.~\ref{fig3}a, contribution due to non-local convection is much
less than the other two quantities in the convective zone, except in
the narrow regions near the boundary of convective zone and in the
overshooting zones. The energy balance in the balk of convective
zone is due to the interplay of the work done by buoyant force
$W_{buo}$ and the viscous dissipation $\rho\epsilon$. Turbulence
retrieves energy from buoyant force, while at the same time it
dissipates energy due to viscosity. The former factor, as the
source, originates primarily from large eddies; while the later one
is happening in the viscous dissipation range of the highest end of
the turbulent spectrum: turbulence gains energy from buoyant force,
which is then cascaded from lowest to higher and higher wave numbers
of turbulent spectrum, and is eventually converted into thermal
energy due to molecular viscosity. It follows from fig.~\ref{fig3}a
that the work done by buoyant force $\rho W_{buo}$ and the viscous
dissipation amount nearly the same but have opposite sign within the
convectively unstable region.

Figs.~\ref{fig3}b and \ref{fig3}c is the enlargements of
fig.~\ref{fig3}a around the two boundaries. Learnt from the two
plots, attention should be paid to the following two points:

\begin{enumerate}

\item In the
overshooting zone, both $\epsilon$ and $W_{buo}$ are negative, while
$dL_r/dM_r >0$. Therefore it is the non-local convective diffusion
that drives overshooting, without which there would be no
overshooting;

\item Before reaching the boundary from the unstable
side, the super-adiabatic temperature gradient has already become
negative, which is more prominent near the lower boundary of the
convective zone (as in fig.~\ref{fig4}). The boundary of convective
zone ($V=0$) is located at $\log P=13.74$ (the vertical dashed
line), but at $\log P\approx 13.34$ still in the unstable zone,
convection is already sub-adiabatic ($\nabla-\nabla_{ad}<0$, see
also fig.~\ref{fig1}). This is distinctly different from the
phenomenological non-local mixing length theories. Although it is
resulted from our special non-local mixing length theory, such
properties of convection ought to be general. This can be proved by
eq.~(\ref{eq14}) of the general dynamic equations of turbulent
convection: The first two terms on the right hand side of
eq.~(\ref{eq14}) are both positive ($x,\,Z>0$). As discussed
already, the third order correlations representing the non-locality
of convection can be neglected compared with the second order terms
in the deep interior of convective zone. Therefore, when $V$ becomes
zero (the fourth term on the right hand side), the super-adiabatic
temperature gradient must be negative ($\nabla-\nabla_{ads}<0$) to
make the equation mathematically right. This proves that
$\nabla-\nabla_{ad}$ must turn negative before $V$ does approaching
the boundary. This nature of non-local convection does not depend on
the kind of non-local convection theory used, which is universally
shared by all theories of non-local convection following
hydrodynamics.

\end{enumerate}

\section{How extended is convective overshooting}\label{sec3}

\begin{figure}
\includegraphics[width=8.4cm]{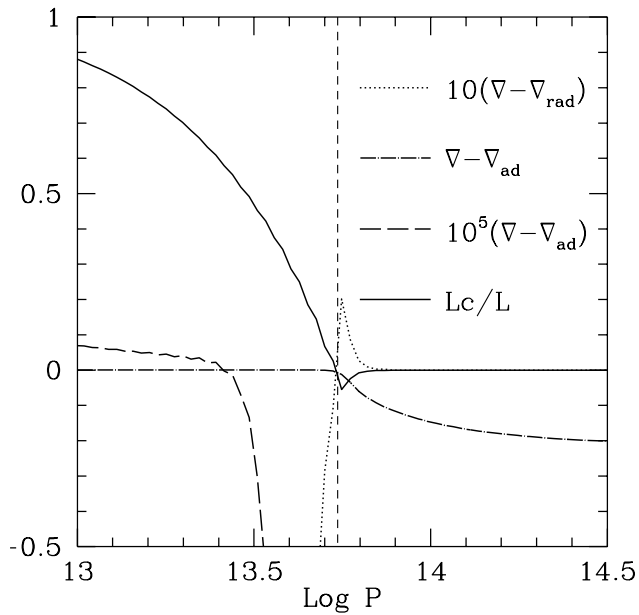}
\caption{The super-adiabatic temperature gradient
$\nabla-\nabla_{ad}$, super-radiative temperature gradient
$\nabla-\nabla_{rad}$ and the fractional convective flux $L_c/L$
versus $\log P$ in the lower convective and overshooting zones for a
non-local convection model of the Sun.}\label{fig4}
\end{figure}

Due to the complexity of non-local convection theory, almost all the
modellings of stellar structure and evolution are still using the
local convection theory. Convective overshooting is defined as the
penetration of convective motion through the classical boundary of
convectively unstable zone into the adjacent stable region. The
extent of overshooting is not the same for different physical
quantities following our dynamic theory of convection, and this
leads to some troubles in understanding and estimation of
overshooting. Followed by the great success of helioseismology,
people are expecting to draw a firm conclusion to the long debated
overshooting distance at the bottom of the solar convective zone
using the helioseismology method. Gough \& Sekii (1993) reported
that they cannot find any definite evidence for the existence of
overshooting in the Sun; while others gave an upper limit of
0.05--0.25H$_P$ (Roxburgh \& Vorontsov 1994, Monteiro et al. 1994,
Christensen-Dalsgaard et al. 1995, Basu \& Antia 1991, Basu 1997).
Such results are understandable. Indeed, what the technique of
helioseismology tests is the (adiabatic) sound speed in the Sun,
while the sound speed is determined by the $P$--$T$ structures. In
the overshooting zone, however, convective flux is negligible. That
is why the overshooting below the bottom of the solar convective
zone has not been detected by helioseismic diagnosis. From
fig.~\ref{fig2}, it is clear that the relative difference in sound
speed between the local and non-local solar models is less than 1\%.
Such tiny differeces were indeed detected by the inversion of
adiabatic sound speed in helioseismology (Basu 1997), however the
observed abrupt increase of the adiabatic sound speed at the bottom
of solar convective zone was not correctly attributed to the
non-local overshooting. In fact, convective flux changes its sign
when crossing the boundary of convective zone, becoming negative in
the overshooting zone ($L_c<0$); in there the radiative flux $L_r$
will be even larger than the total flux of the Sun ($L_\odot$). In
the overshooting zone, the temperature gradient will overtake the
radiative counterpart $\nabla>\nabla_{rad}$ (see fig.~\ref{fig4}).
As a result, the temperature at the bottom of solar convective zone
will rise up, just as what has been detected by helioseismology
technique.

\begin{figure}
\includegraphics[width=8.4cm]{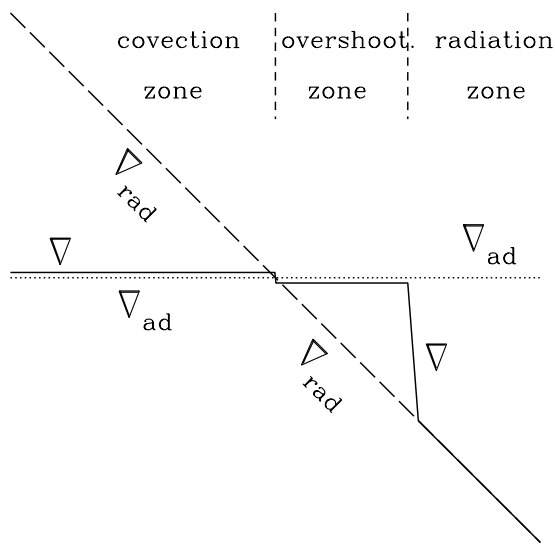}
\caption{A sketch of the lower convective and overshooting zones in
the usual phenomenological non-local mixing length theory (Monteiro
et al. 2000).}\label{fig5}
\end{figure}

\begin{figure}
\includegraphics[width=8.4cm]{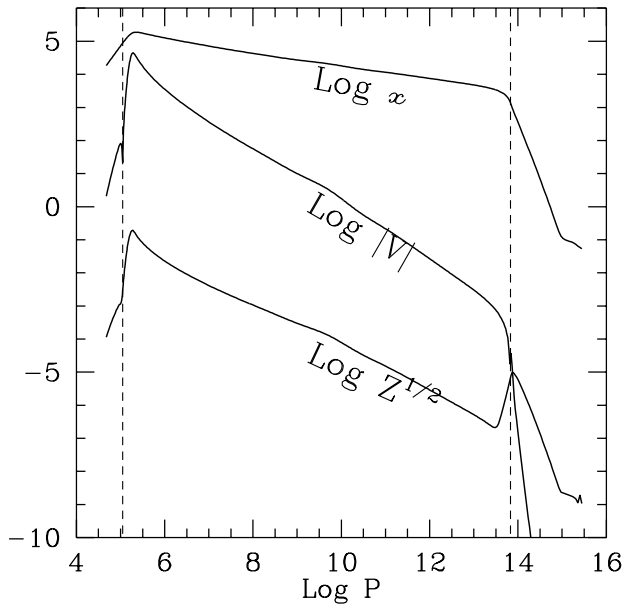}
\caption{The auto- and cross-correlations of turbulent velocity and
temperature $x$, $Z^{1/2}$ and $V$ versus depth ($\log P$) for a
non-local convection model of the Sun. The vertical dashed lines
indicate the upper and lower boundaries of the convective
zone.}\label{fig6}
\end{figure}

Gough \& Sekii (1993) measured the extension of overshooting at the
bottom of solar convective zone following a picture of the
overshooting zone made by the phenomenological non-local mixing
length theory, which is illustrated in fig.~\ref{fig5}. In the
unstable zone, the temperature gradient is slightly higher than the
adiabatic one, while being slightly lower in the overshooting zone.
After a distance of an overshooting length $d_{ov}$, the temperature
gradient switches suddenly to the radiative from adiabatic, making a
discontinuity in temperature gradient at the bottom of the
overshooting zone. The jump size of the temperature gradient is
proportional to the overshooting distance $d_{ov}$. Such a
discontinuity in temperature gradient is exactly what Gough \& Sekii
used to detect the overshooting distance $d_{ov}$. It is the
implicit assumption of full (either positive or negative)
correlation between turbulent velocity and temperature fluctuations
that makes the misunderstandings of overshooting zone in the
non-local mixing length theory (Xiong 1985, Petrovay \& Marik 1995).
In reality, however, the turbulent velocity-temperature correlation
decreases very quickly and approaches zero near the lower boundary
of the solar convective zone where convective energy transfer is
very efficient, as demonstrated in fig.~\ref{fig1}. Therefore, there
is no similarity in the structure overshooting zone between the
phenomenological non-local mixing length theory and the dynamic
theory of non-local convection. In our view, the temperature
gradient has already been smaller than the adiabatic one
($\nabla<\nabla_{ad}<\nabla_{rad}$) before reaching the lower
boundary of convective zone. The convective flux becomes negative
passing through the boundary, therefore the temperature gradient
$\nabla$ is smaller than the adiabatic temperature gradient
$\nabla_{ad}$ and higher than the radiative one $\nabla_{rad}$
($\nabla_{ad}<\nabla<\nabla_{rad}$). The temperature gradient
changes continuously instead of abruptly from $\nabla_{ad}$ to
$\nabla_{rad}$. The structure of the overshooting zone in our
dynamic theory of non-local convection is shown in fig.~\ref{fig4}.
In the overshooting zone under the convective zone, there is a
narrow ($\sim 0.25H_P$) and weakly super-radiative region. Actually,
the overshooting zone is nearly radiative rather than nearly
adiabatic. Therefore, it is not a surprise why Gough \& Sekii (1993)
could not find any firm evidence for the existence of the
overshooting under the bottom of solar convective zone. We can
further justify that the methods based on stellar thermal ($P$-$T$)
structure will all under-estimate the true overshooting distance. In
the lower overshooting, the turbulent velocity-temperature correlation is
very small, and convective energy transfer in there is negligible.
The overshooting distance detected by convective energy flux will be
far smaller than that is represented by the turbulent velocity and
temperature fields. It follows from fig.~\ref{fig6} that, in the
overshooting zones either at the surface or bottom of the solar
convective zone, the overshooting distances of turbulent velocity
and temperature are both very extended, their e-folding lengths
of overshooting are given in table~\ref{tab1}. Our
theoretical e-folding length agrees fairly well with those derived
from observations of the solar granular field (Keil \& Danfield
1978, Nesis \& mattig 1989, Komm, Mattig \& Nesis 1991).

\begin{table}
\scriptsize \caption{The e-folding lengths.}\label{tab1}
\begin{tabular}{ccccccccc}
 \hline
 & \multicolumn{3}{c}{upper oversh. zone}
 &\multicolumn{3}{c}{lower oversh. zone} & & \\
 M/M$_\odot$ &   x  &  z  &  V  &  x  &  z  &  V  &  Li &  Dcut\\
\hline
 0.800 & 0.47 & 0.36 & 0.20 &  0.25 & 0.25 & 0.080 & 0.26 &  0.30\\
 0.850 & 0.48 & 0.36 & 0.21 &  0.25 & 0.25 & 0.081 & 0.36 &  0.38\\
 0.900 & 0.50 & 0.35 & 0.20 &  0.25 & 0.25 & 0.082 & 0.42 &  0.54\\
 0.925 & 0.50 & 0.36 & 0.21 &  0.25 & 0.25 & 0.069 & 0.50 &  0.58\\
 0.950 & 0.50 & 0.33 & 0.20 &  0.26 & 0.25 & 0.076 & 0.60 &  0.69\\
 0.975 & 0.49 & 0.33 & 0.20 &  0.31 & 0.31 & 0.104 & 0.67 &  0.74\\
 1.000 & 0.50 & 0.32 & 0.20 &  0.29 & 0.31 & 0.093 & 0.85 &  0.91\\
 1.025 & 0.63 & 0.30 & 0.21 &  0.29 & 0.29 & 0.096 & 1.05 &  1.11\\
 1.050 & 0.52 & 0.30 & 0.19 &  0.30 & 0.36 & 0.092 & 1.25 &  1.33\\
 1.075 & 0.60 & 0.30 & 0.21 &  0.38 & 0.38 & 0.114 & 1.64 &  1.69\\
 \hline
\end{tabular}
\end{table}

The overshooting distance in terms of stellar evolution is the
extension of the non-local convective mixing of chemical elements.
Obviously, it is neither that of convective energy transfer nor
those of turbulent velocity and temperature fluctuations.
Calculations of massive star evolution under our complete non-local
theory of convection shown that the non-local convective mixing
overshoots a very extended distance (Xiong 1986).

\begin{figure*}
\includegraphics[width=16.8cm]{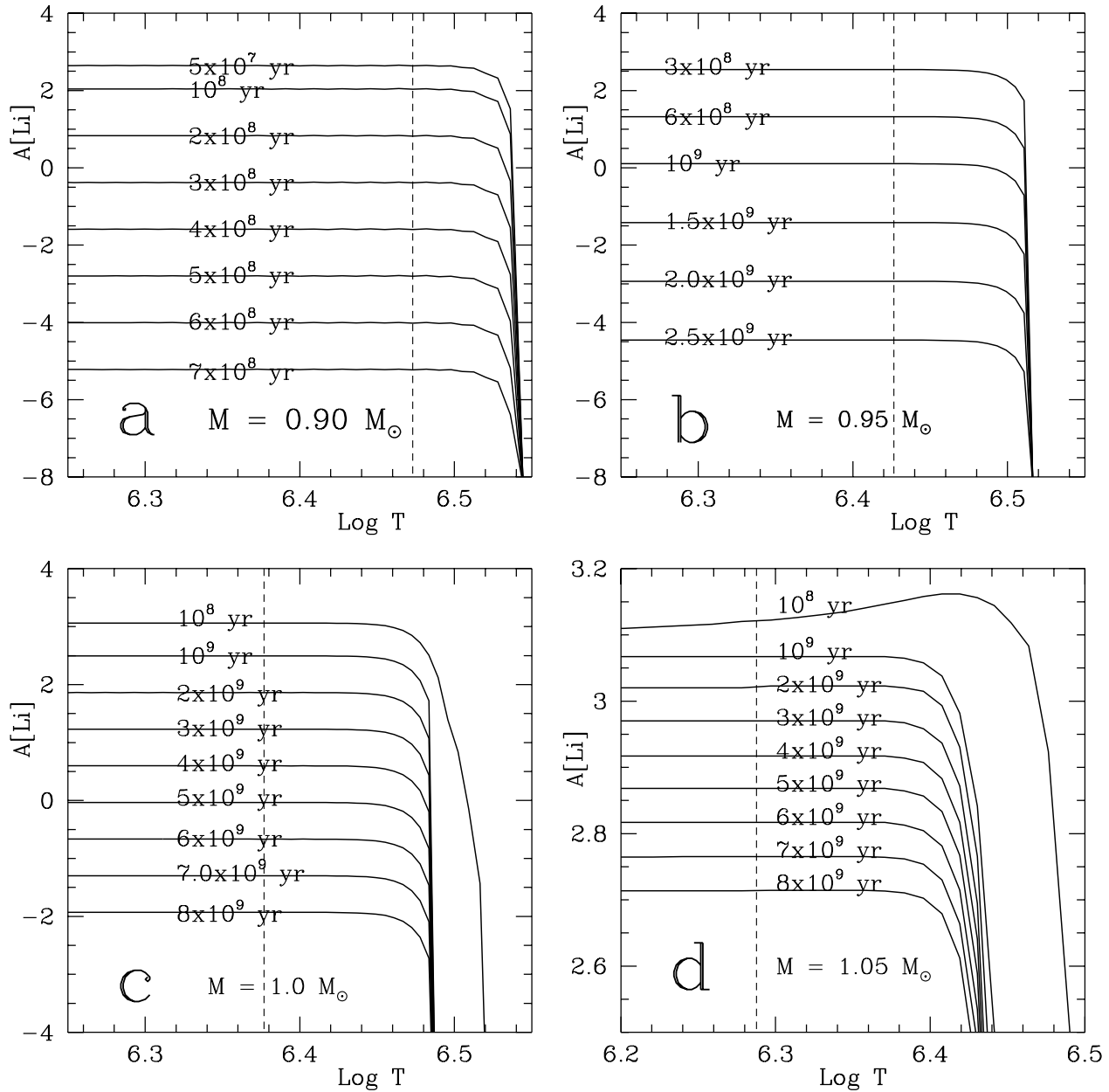}
\caption{The Lithium abundance versus depth ($\log T$) and age
(labeled on the curves) for main sequence stars. The dashed vertical
line indicates the location of the lower convective boundary.
Evolution is not considered in the calculations. a). M=0.90M$_\odot$; b).
M=0.95M$_\odot$, c). M=1.0M$_\odot$ and d).
M=1.05M$_\odot$.}\label{fig7}
\end{figure*}

Although the overshooting at the bottom of the solar convective zone
cannot be observed directly, we fortunately have another excellent
indicator, which is the Lithium abundance of the Sun and solar type
stars, for the extension of overshooting. No matter how disputed the
mechanism of Lithium depletions in the atmospheres of solar type
stars is, it can provide, at least, an upper limit for the extension
of overshooting zone in these stars. As it is well known, $^7$Li
gets burnt due to reaction $^7Li(P,\alpha)^4He$ at a temperature of
$2.5\times 10^6$K. The depth of the solar convective zone, as given
by helioseismology, is $r_c/R_\odot\approx 0.713$, and the
temperature at the bottom of the zone is $T_c\approx 2.26\times
10^6$K (Basu \& Antia 1997, Christensen-Delsgaard et al. 1991). This
temperature is not high enough to burn Lithium, without overshooting
(bringing Lithium deeper to higher temperatures) there will be no
depletion of Lithium in the Sun. It is the overshooting that brings
Lithium to the burning region at a higher temperature, and causes
depletion. Figs.~\ref{fig7}a--\ref{fig7}d show the Lithium abundance
depletions due to overshooting for M=0.90,0.95,1.00 and
1.05M$_\odot$ stellar models as functions of depth ($\log P$), when
the effect of evolution is not considered. The dashed lines in the
plots indicate the boundary of convective zone. As from
fig.~\ref{fig7}c, there is a gradually accelerating reduction of
Lithium abundance in the overshooting zone. In the upper part of the
overshooting zone, the mixing caused by non-local convection is very
efficient, the abundance keeps the same as in the convectively
unstable zone for about 0.4 H$_P$ in length downwards. It is then
followed by a partially mixed region where Lithium abundance is
reduced quicker toward the center and vanishes suddenly, such a
partial mixing region is about 0.5H$_P$ in depth. Even deeper is the
non-mixing zone. If the overshooting distance is taken as the e-folding
length of the abundance from the bottom of the convective zone, it
reads about 0.83H$_P$; otherwise if we count the deeppest bottom of the
mixing process, it reads about 0.9H$_P$ for the solar model.

The equation for the conservation of Lithium abundance can be
written as,

\bq {1\over C}\frac{\partial C}{\partial t}=-{1\over
C}\frac{\partial}{\partial M_r}\left(4\pi r^2\rho
U\right)-q,\label{eq35}\eq

\begin{figure*}
\includegraphics[width=16.8cm]{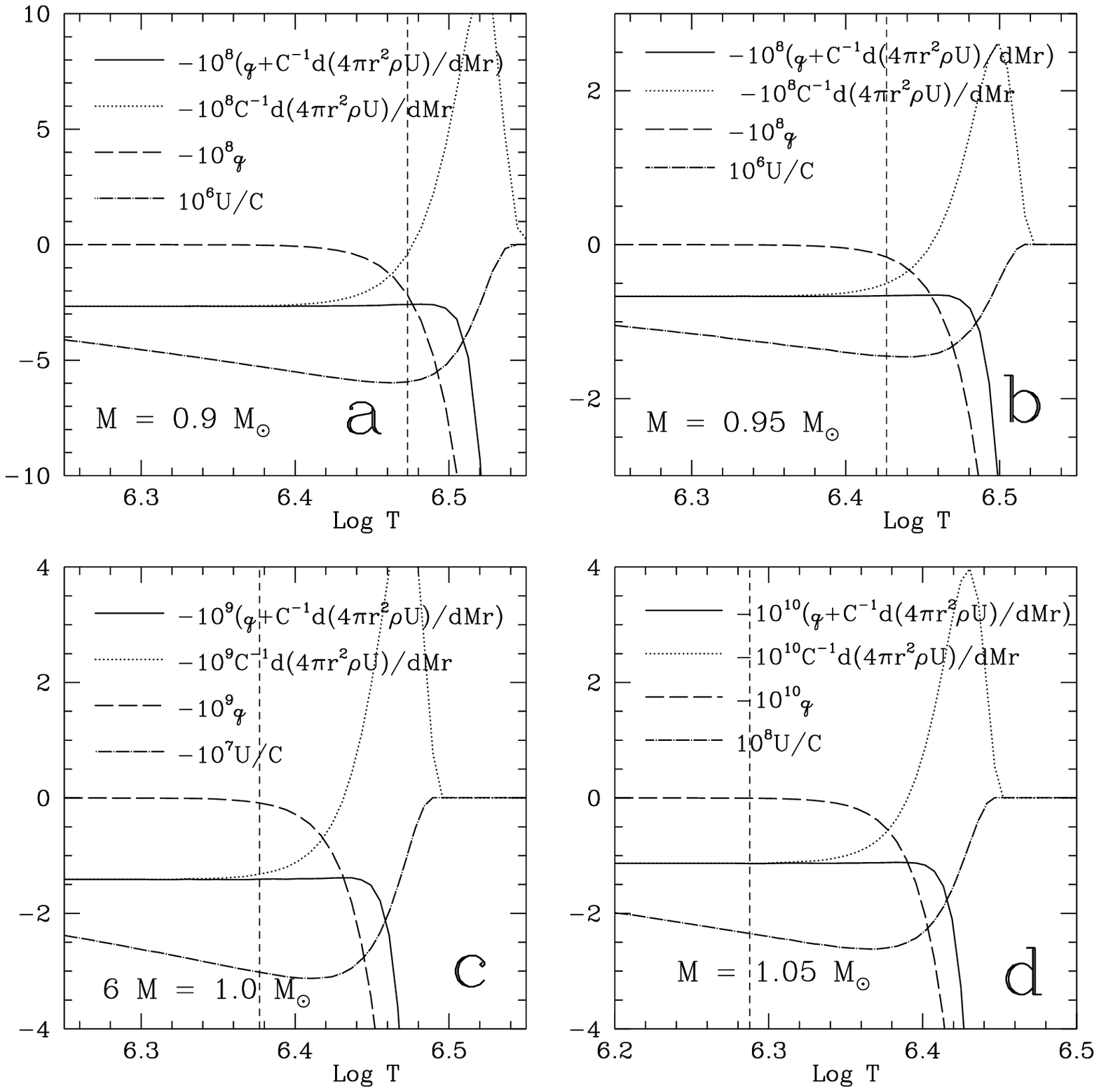}
\caption{$U/C$ ($sec^{-1}$), burning rate of Lithium $^7$Li
(yr$^{-1}$) $q$, change rate of convective diffusion for Lithium
$\eta=-{1\over C}\frac{\partial}{\partial M_r}\left(4\pi r^2\rho
U\right)$(yr$^{-1}$) and $\eta-q$ versus depth. The dashed line
indicates the lower boundary of the convective zone. a).
M=0.90M$_\odot$; b). M=0.95M$_\odot$, c). M=1.0M$_\odot$ and d).
M=1.05M$_\odot$. }\label{fig8}
\end{figure*}
\begin{figure}
\includegraphics[width=8.4cm]{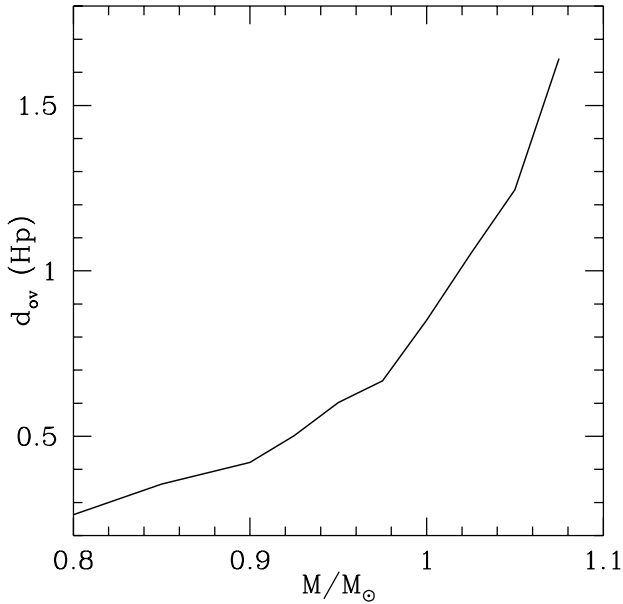}
\caption{The overshooting distance as a function of stellar mass, as
derived from Lithium depletion in solar type stars.}\label{fig9}
\end{figure}


where $C$ is the Lithium abundance by mass, $q$ is burning rate of
Lithium, and U is the correlation of the radial component of
turbulent velocity $w'_r$ and the turbulent fluctuations of Lithium
abundance defined as,

\bq U=\overline{w'_rC},\label{eq36}\eq

therefore $4\pi r^2\rho U$ is the total flux of convective mixing of
Lithium passing through the sphere of radius $r$, and the first term
on the right hand side of eq.~(\ref{eq35}) is the rate of variation
of Lithium abundance due to non-local convective mixing. It follows
from fig.~\ref{fig6} that, in the overshooting zone, turbulent
velocity decreases nearly exponentially. The non-mixed region does
not mean there is no mixing at all. In our picture, mixing is always
there, the only difference is quantity of mixing. When the non-local
mixing timescale $\tau_{mix}$ becomes much longer then the nuclear
timescale $\tau_{nuc}$ of depletion, i.e. the non-local mixing
cannot feed fresh Lithium into the burning zone, Lithium abundance
vanished abruptly. Figs~\ref{fig8}a--d give these two terms on the
right hand side of eq.~(\ref{eq35}) and their sum (the depletion
rate of Lithium) versus depth near the lower boundary of convective
zones for 0.90, 0.95, 1.00 and 1.05M$_\odot$ main sequence stellar
models respectively. The unit for these quantities are all
$yr^{-1}$. The vertical dashed line locates the lower boundaries in
these models. As clearly shown in fig.~\ref{fig8}, the depletion of
Lithium starts at $T\sim 2.5\times 10^6K$, and it goes up very
quickly as temperature increases (being proportional to the 21st
power of temperature). In the whole envelopes of these models,
$U<0$, this means that convection keeps feeding Lithium from outer
into inner layers in order to supply the depletion at the burning
zone. In the convectively unstable zone (where $q$ is extremely
small), convective mixing is very efficient, $d\left(4\pi r^2\rho
U\right)/dMr$ is nearly constant (see fig.~\ref{fig8}). Towards the
deep interior, the nuclear burning rate $q$ goes up abruptly. In the
upper part of the overshooting zone, mixing due to convective
overshooting is still efficient enough to compensate the depletion
due to nuclear burning, therefore the Lithium abundance profile is
still horizontal (see figs.~\ref{fig7} and \ref{fig8}). At the lower
part of the overshooting zone, however, the convective overshooting
mixing (the dotted lines in fig.~\ref{fig8}) can no longer support
the balance between mixing and depletion (long dashed lines in
fig.~\ref{fig8}). $-{1\over C}d(4\pi\rho U)/dM_r-q$ decrease
abruptly towards the center, corresponding to Lithium abundance
dropping off in fig.~\ref{fig7}. This means the boundary of
overshooting zone is reached at this place. For the M=0.90M$_\odot$
star, convective zone is very deep with its bottom already at the
burning region of Lithium, this may leads to a shallower
overshooting zone. Toward higher masses, convection becomes
shallower, and the bottom of the zone goes farther away from the
region of burning (see figs.~\ref{fig8}c--d), the overshooting zone
is then becoming more extended. Fig.~\ref{fig9} presents the
overshooting distance $d_{ov}$ measured by Lithium depletion as a
function of stellar mass, from which it is clear that the
overshooting distance $d_{ov}$ increases as stellar mass increases,
going from 0.26H$_P$ for M=0.80M$_\odot$ to 1.65H$_P$ for
M=1.075M$_\odot$. The overshooting distance defined by dropping off
of Lithium abundance by a factor of $e$ is shown in the 8th column
of table~\ref{tab1}, while that defined by the distance from the
boundary of convective zone to where Lithium becomes zero is given
in the last column of table~\ref{tab1}. It is clearly from
figs.~\ref{fig7}a--d that Lithium abundance vanishes very quickly
after the e-folding depletion, the distance between them is less
than 0.1H$_P$. Completely different from the Lithium abundance
profile in fig.~\ref{fig7}, turbulent velocity ($x$), temperature
fluctuation ($Z^{1/2}$) and velocity--temperature correlation ($V$)
decrease exponentially with depth in the overshooting zone. The
e-folding distances determined from the curves are given in the
2nd--4th colums (for the upper part of overshooting zone), and
4th--7th columns (the lower part of overshooting zone). The
e-folding distances given by turbulent velocity and temperature
fields are very close to the analytic asymptote in our theory (Xiong
1989b). They hardly change with stellar mass, and are rather
different from the e-folding distances defined by Lithium depletion.
The upper and lower overshooting zones are a bit different in terms
of these e-folding distances, this is due to the fact that, in the
overshooting zone above the convective zone, gas density is low so
that the effective Peclet number $P_e\ll 1$, and convective energy
transfer is inefficient; while on the contrary, in the overshooting
zone attach to the bottom of the convective zone, $P_e\gg 1$, and
convection is highly efficient. These arguments mark the distinct
properties of the velocity--temperature correlation in the upper and
lower overshooting zones: in the surface overshooting zone,
$R_{VT}\sim -1$, while in the bottom overshooting zone, $R_{VT}\sim
-0.0$. When plotting fig.~\ref{fig1}, $R_{VT}$ has been magnified in
order to show the details. In fact, we should have $-10^3<R_{VT}<0$
in the lower overshooting zone, so that it is completely buried in
the solid line of ($L_c/L$).

\section{Summary and discussions}\label{sec4}

Detailed discussions on the definition of the boundary of convective
zone, and the distance of convective overshooting are presented in
this paper. The main results can be summarized in the following:

\begin{enumerate}

\item Choosing the place where the convective flux (or equivalently
the turbulent velocity--temperature correlation) changes it sign as
the boundary is the most proper and convenient. Where the convective
flux is greater than zero is convectively unstable zone, while it is
the overshooting zone when the convective flux is smaller than zero.
The convective zone defined as such not only makes the local the
non-local convection models with the same depth of convection zone
to have similar structures, but also to have very clear physical
meanings: convective zone is the buoyant force driving zone for
convective motion, while the overshooting zone is the dissipation
zone against convective motion, which can only be supported by
non-local convective diffusion;

\item It is not quite right to talk about a general overshooting
distance for stellar convection. The distance of overshooting is
different for different physical quantities. The effect of
overshooting for convective energy transfer, for instance, is not
important. However, the overshooting distances of turbulent velocity
and temperature fluctuations are quite extended, the e-folding
lengths can reach 0.25--0.5H$_P$. The overshooting distance in term
of stellar evolution is the extent of convective mixing of chemical
elements. From the example set by the depletion of Lithium in solar
type stars, it is found that convective mixing of matter happens in
stellar evolutionary (nuclear) timescales, and is very efficient.
Very extended and weak overshooting can still induce fairly
efficient mixing in a very long timescale of evolution. Therefore we
anticipate a very extended overshooting for mixing of matters, the
e-folding lengths of which is generally larger than that of
turbulent velocity and temperature. The one in massive stellar
model, for instance, can reach 1 pressure scale height (Xiong 1986).

\item The problem of Lithium depletion in solar type stars is a
special case of overshooting mixing that moves Lithium to the
burning region from surface. Under such circumstance, overshooting
distance depends on the location of convective zone relative to that
of Lithium burning region. Obviously, such conclusion cannot be
simply generalized to the case of the core nuclear reaction. For
Lithium depletion in solar type stars, the convective zone becomes
shallower for higher masses, and Lithium depletion becomes slower
(longer timescale), and this makes the extended overshooting tail to
be efficient in mixing, and the mixing range to become more
extended. As a result, the overshooting distance detected by Lithium
depletion in solar type stars becomes larger for increasing stellar
mass. In fact, this indicates that mixing of matter becomes more
extended for increasing nuclear burning timescales.

\end{enumerate}

\section*{Acknowledgements}
The Chinese National Natural Science Foundation (CNNSF) is
acknowledged for support through grants 10573022, 10173013, 10273021
and 10333060.


\begin{thebibliography}{99}
\bibitem{BA94}Basu, S. \& Antia, H.M., 1994, MNRAS, 269, 1137
\bibitem{BA97}Basu, S. \& Antia,H.M., 1997, MNRAS, 287, 189
\bibitem{Basu97}Basu, S., 1997, in Proc. IAU Symp. 181, Sounding solar and
stellar interiors, eds. J. Provost, F-X Schmider (Kluwer,
Dordrecht), p. 137
\bibitem{BV58}B\"ohm-Vitense, E., 1958, Astrophysik.
46, 108
\bibitem{BBC81}Bressan, A., Bertelli, G. \& Chiosi, C., 1981, A\&A,
102, 25
\bibitem{can93}Canuto, V.M., 1993, ApJ, 416, 331
\bibitem{CD98}Canuto, V.M. \& Dubovikov, M., 1998, ApJ, 493, 834
\bibitem{CS70}Chiosi, C. \& Summa, C., 1970, ApSS, 8, 478
\bibitem{CMT95}Christensen-Dalsgaard, J., Monteiro, M.J.P.F.G., Thompson, M.J.,
1995, MNRAS, 276, 283
\bibitem{CGT91}Christensen-Dalsgaard, J., Gough, D.O. \& Thompson,
M.J., 1991, ApJ, 378, 413
\bibitem{GS93}Gough, D.O. \& Sekii, T. 1992, in ASP Conf. Ser. Vol
42, ed. T.M. Brown, p.117
\bibitem{GBA93}Grossman, S.A., Narayan, R. \& Arnett, D., 1993, ApJ,
407, 284
\bibitem{Gross96}Grossman, S.A., 1996, MNRAS, 279, 305
\bibitem{KC78}Keil, S.L. \& Canfield, R.C., 1978, A\&A, 70, 169
\bibitem{KMN91}Komm, R., Mattig, R.W. \& Neiss, A., 1991, A\&A, 243,
251
\bibitem{Kuh86}Kuhfuss, R., 1986, A\&A, 160, 116
\bibitem{KPK03}Kupka, F., 2003, in Modelling of Stellar Atmosphere,
IAU Symp. 210, eds N. Piskunov, W.W. Weiss \& D.G. Gray, p.143 (Pub.
ASP)
\bibitem{LNS62}Leighton, R.B., Neyes, R.W. \& Simon, G.W., 1962,
ApJ, 135, 474
\bibitem{MCT94}Monteiro, M.J.P.F.G., Christensen-Dalsgaard, J. \&
Thompson, M.J., 1994, A\&A, 283, 247
\bibitem{MCT00}Monteiro, M.J.P.F.G., Christensen-Dalsgaard, J. \&
Thompson, M.J., 2000, MNRAS, 316, 165
\bibitem{NM89}Nesis, A. \& Mattig, W., 1989, A\&A, 221, 130
\bibitem{RV94}Roxburgh, I.W. \& Vorontsov, S.V., 1994, MNRAS, 268,
880
\bibitem{SBGC04}Salucci, G., Bertello, L., Gavallini, F. Ceppatelli, G. et al., 2004, MNRAS, 285, 322
\bibitem{SH58}Schwarzschild, M. \& Harm, R., 1958, ApJ, 128,348
\bibitem{Spie63}Spiegel, E.A., 1963, 216
\bibitem{Sto70}Stothers, R., 1970, MNRAS, 151, 65
\bibitem{TM73}Travis, L.D. \& Matsushima, S., 1973, ApJ, 186, 975
\bibitem{Ulr70}Ulrich, R.K., 1970, ApSS, 7, 183
\bibitem{UKX85}Unno, W., Kondo, M. \& Xiong, D.R. 1985, PASj, 37,
235
\bibitem{X77}Xiong, D.R., 1977, AcASn, 18, 86
\bibitem{X78}Xiong, D.R., 1978, ChA, 2, 118
\bibitem{X81a}Xiong, D.R., 1981a, SciSn, 23, 1139
\bibitem{X81b}Xiong, D.R., 1981b, AcASn, 22, 356
\bibitem{X82}Xiong, D.R., 1982, ChA, 6, 43
\bibitem{X85}Xiong, D.R., 1985, A\&A, 150, 133
\bibitem{X86}Xiong, D.R., 1986, A\&A, 167, 239
\bibitem{X89a}Xiong, D.R., 1989a, A\&A, 209, 126
\bibitem{X89b}Xiong, D.R., 1989b, A\&A, 213, 176
\bibitem{XC92}Xiong, D.R. \& Cheng, Q.L., 1992, A\&A, 254, 362
\bibitem{XDC98}Xiong, D.R., Deng, L. \& Cheng, Q.L., 1998, ApJ,
499, 355
\bibitem{XCD98}Xiong, D.R., Cheng, Q.L. \& Deng, L., 1998, ApJ,
500, 449
\bibitem{XD2001a}Xiong, D.R. \& Deng, L., 2001, MNRAS, 324, 243
\bibitem{XD2007}Xiong, D.R. \& Deng, L., 2007, MNRAS, in press

\end{thebibliography}
\end{document}